\documentclass[prl,twocolumn,floatfix,showpacs ]{revtex4-1}
\UseRawInputEncoding
\usepackage{amsmath}
\usepackage{amssymb}
\usepackage{graphicx}
\usepackage{dcolumn}
\usepackage{bm}
\usepackage[colorlinks=true,linkcolor=blue,anchorcolor=blue, citecolor=cyan,urlcolor=cyan]{hyperref}
\usepackage[mathlines]{lineno}
\usepackage{ulem}
\usepackage{epstopdf}
\usepackage{placeins}
\begin{document}
\title{Weak valley-layer coupling and valley polarization in  centrosymmetric   $\mathrm{FeCl_2}$ monolayer}
\author{San-Dong Guo}
\email{sandongyuwang@163.com}
\affiliation{School of Electronic Engineering, Xi'an University of Posts and Telecommunications, Xi'an 710121, China}
\author{Liguo Zhang}
\affiliation{School of Electronic Engineering, Xi'an University of Posts and Telecommunications, Xi'an 710121, China}
\author{Xiao-Shu Guo}
\affiliation{School of Electronic Engineering, Xi'an University of Posts and Telecommunications, Xi'an 710121, China}
\author{ Gangqiang Zhu}
\affiliation{School of Physics and Electronic Information, Shaanxi Normal University, Xi'an 716000, Shaanxi, China}
\begin{abstract}
Using the valley degree of freedom as a carrier of information for storage and processing, valley polarization plays a crucial role.
A variety of mechanisms for valley polarization have been proposed, among which the valley-layer coupling mechanism involves the induction of valley polarization by an out-of-plane electric field. Here, through first-principles calculations, it is found that  the weak valley-layer coupling can exist  in centrosymmetric $\mathrm{FeCl_2}$ monolayer. It is crucial to note that valley-layer coupling only occurs with out-of-plane magnetization and vanishes with in-plane magnetization. Compared to monolayers with strong valley-layer coupling, $\mathrm{FeCl_2}$ requires an extremely strong electric field to achieve the same magnitude of valley splitting. Valley polarization switching can be achieved by manipulating the directions of magnetization and electric field. Reversing only one of these directions switches the valley polarization, whereas reversing both simultaneously leaves it unchanged.
Moreover, the simply stacked bilayer $\mathrm{FeCl_2}$, as a $PT$-antiferromagnet, can spontaneously achieve valley polarization without an external electric field, highlighting its potential for miniaturization, ultradensity, and ultrafast performance. Our work provides guidelines for identifying materials with weak valley-layer coupling, and further enables the regulation of valley polarization through electric field and stacking engineering.

\end{abstract}
\maketitle
\textcolor[rgb]{0.00,0.00,1.00}{\textbf{Introduction.---}}
In valleytronic materials, the valley degree of freedom, distinct from charge and spin, emerges as an information carrier, enabling the processing of data and the execution of logic operations\cite{q1,q2,q4}.
Valleys, which represent the energy extrema of bands, exhibit remarkable robustness against smooth deformations and low-energy phonons due to their large separation in momentum space. The burgeoning discovery of diverse two-dimensional (2D) materials has invigorated the field of valley physics. A prime example is the family of 2D nonmagnetic transition metal dichalcogenides (TMDs), which feature a pair of degenerate yet inequivalent valleys, -K and K, in reciprocal space\cite{q8-1,q8-2,q8-3,q9-1,q9-2,q9-3}. However, the presence of time-reversal symmetry ($T$) in these nonmagnetic TMDs results in the degeneracy of the -K and K valleys, thereby precluding valley polarization. While external magnetic fields, proximity effects, magnetic doping, and light excitation have been employed to induce valley polarization, these methods often suffer from significant drawbacks,  including minimal valley splitting, potential disruption of the crystal structure, and limited carrier lifetimes\cite{v5-1,q9-3,q9-2,v10-1}.

The 2D  hexagonal ferromagnetic (FM)  materials, which naturally break $T$ symmetry, offer an ideal platform for manipulating valley polarization. When these magnetic materials also break space inversion symmetry ($P$), valley polarization between the -K and K valleys can be induced. Building on this foundation, the concept of the "ferrovalley"  is first proposed\cite{q10}. Ferrovalley materials exhibit spontaneous valley polarization, thereby enabling the realization of the anomalous valley Hall effect (AVHE)\cite{q11,q12,q13,q13-1,q14,q14-1,q15,q16,q18}. Zero-net-magnetization magnets offer several advantages over ferromagnets for valleytronic devices, which  include the ability to achieve higher storage densities, enhanced robustness against external magnetic fields, and ultrafast writing speeds\cite{k1,k2}.
Some possible design strategies to achieve valley polarization  in 2D  zero-net-magnetization magnets, for example $PT$-antiferromagnet (the joint symmetry ($PT$) of $P$ and $T$), altermagnet and fully-compensated ferrimagnet, have been proposed, which has broadened the field of valleytronics\cite{gsd0,gsd1,gsd2,k4,k5,k6,k7,k8,k9,k10}.

\begin{figure}[t]
    \centering
    \includegraphics[width=0.40\textwidth]{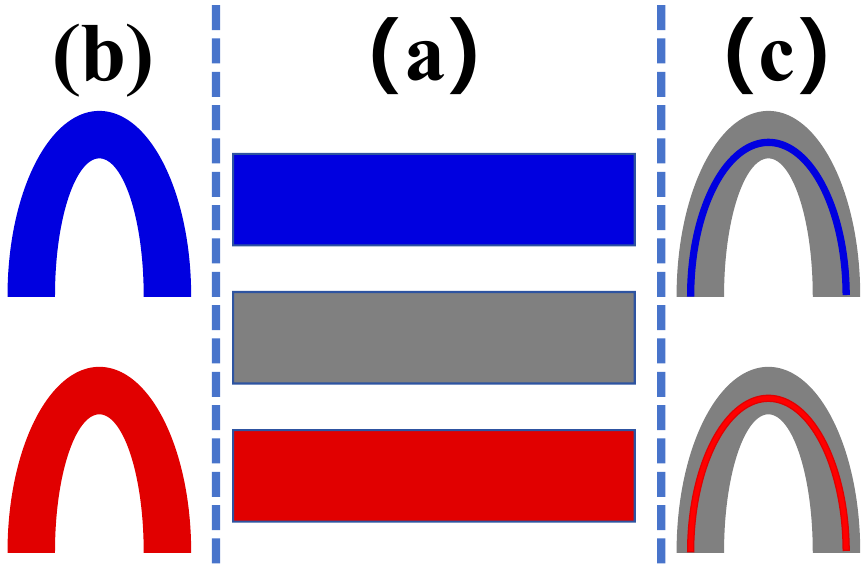}
     \caption{(Color online) (a):the middle layer (gray) separates the
 unit cell into  upper layer (blue)  and  lower layer (red); (b): the two valleys are mainly contributed by the upper layer and the lower layer, respectively, resulting in strong valley-layer coupling; (c): the two valleys are both primarily contributed by the middle layer, and also have minor contributions from the upper and lower layers, respectively,  giving rise to weak valley-layer coupling. In (b) and (c), the colors are associated with the layers, and the thickness of the lines indicates the weight of the layer contributions.}\label{sy}
\end{figure}

However, electric control of valley polarization is highly desirable for device applications due to its advantages in compactness, power efficiency, and compatibility with existing semiconductor technologies. Unfortunately, an electric field alone does not break $T$ symmetry, which is not conducive to the manipulation of valley polarization. Nevertheless, if a system with three sectors (\autoref{sy} (a)) exhibits valley-layer coupling, an out-of-plane electric field can induce valley polarization through the creation of a potential difference\cite{k11,k12,k13}.
In prior research, a strong coupling between valley and layer has been observed, where the two valleys are predominantly influenced by the contributions from the upper and lower layers (\autoref{sy} (b)), respectively. Beyond this scenario, a weak coupling between valley and layer may also exist. In this case, both valleys are primarily sourced from the middle layer, while the upper and lower layers each provide minor contributions to the two valleys (\autoref{sy} (c)).
Compared with materials having strong valley-layer coupling, a significantly stronger electric field is required to induce the same magnitude of valley splitting in materials with weak valley-layer coupling.

Here, it is found that  the weak valley-layer coupling can exist  in centrosymmetric $\mathrm{FeCl_2}$  with out-of-plane magnetization.
The electric field can indeed induce valley splitting in $\mathrm{FeCl_2}$, which is consistent with the exploitable results\cite{q1811}.
Compared with the existing results\cite{q1811}, we have provided a more intuitive and comprehensive explanation of the electric field-induced valley polarization in $\mathrm{FeCl_2}$, and have proposed a simple bilayer-stacked $\mathrm{FeCl_2}$ with $P$ lattice symmetry, which can spontaneously exhibit valley polarization without the need for an external electric field.

\textcolor[rgb]{0.00,0.00,1.00}{\textbf{Computational detail.---}}
  The spin-polarized first-principles calculations within density functional theory (DFT)\cite{1}are performed using the Vienna Ab Initio Simulation Package (VASP)\cite{pv1,pv2,pv3}. The kinetic energy cutoff  of 500 eV,  total energy  convergence criterion of  $10^{-8}$ eV, and  force convergence criterion of 0.001 $\mathrm{eV.{\AA}^{-1}}$ are set to obtain reliable results. We use  Perdew-Burke-Ernzerhof generalized gradient approximation (PBE-GGA)\cite{pbe} as the exchange-correlation functional to optimize the lattice parameters and conduct various energy comparisons. To account for the localized nature of Fe-3$d$ orbitals, a Hubbard correction $U_{eff}$=4 eV\cite{fecl}  is applied using the rotationally invariant approach proposed by Dudarev et al\cite{du} to investigate electronic structures. The  spin-orbital coupling (SOC) is incorporated for investigation of electronic structure and magnetic anisotropy energy (MAE).
 A vacuum space of more than 18 $\mathrm{{\AA}}$ along the $z$-direction is introduced to prevent interactions between neighboring slabs. The Brillouin zone (BZ) is sampled using a 21$\times$21$\times$1 Monkhorst-Pack $k$-point meshes. The magnetic orientation is determined  by calculating  MAE: $E_{MAE}=E^{||}_{SOC}-E^{\perp}_{SOC}$, where $||$ and $\perp$  mean that spins lie in
the plane and out-of-plane.

\begin{figure}
 \includegraphics[width=0.45\textwidth]{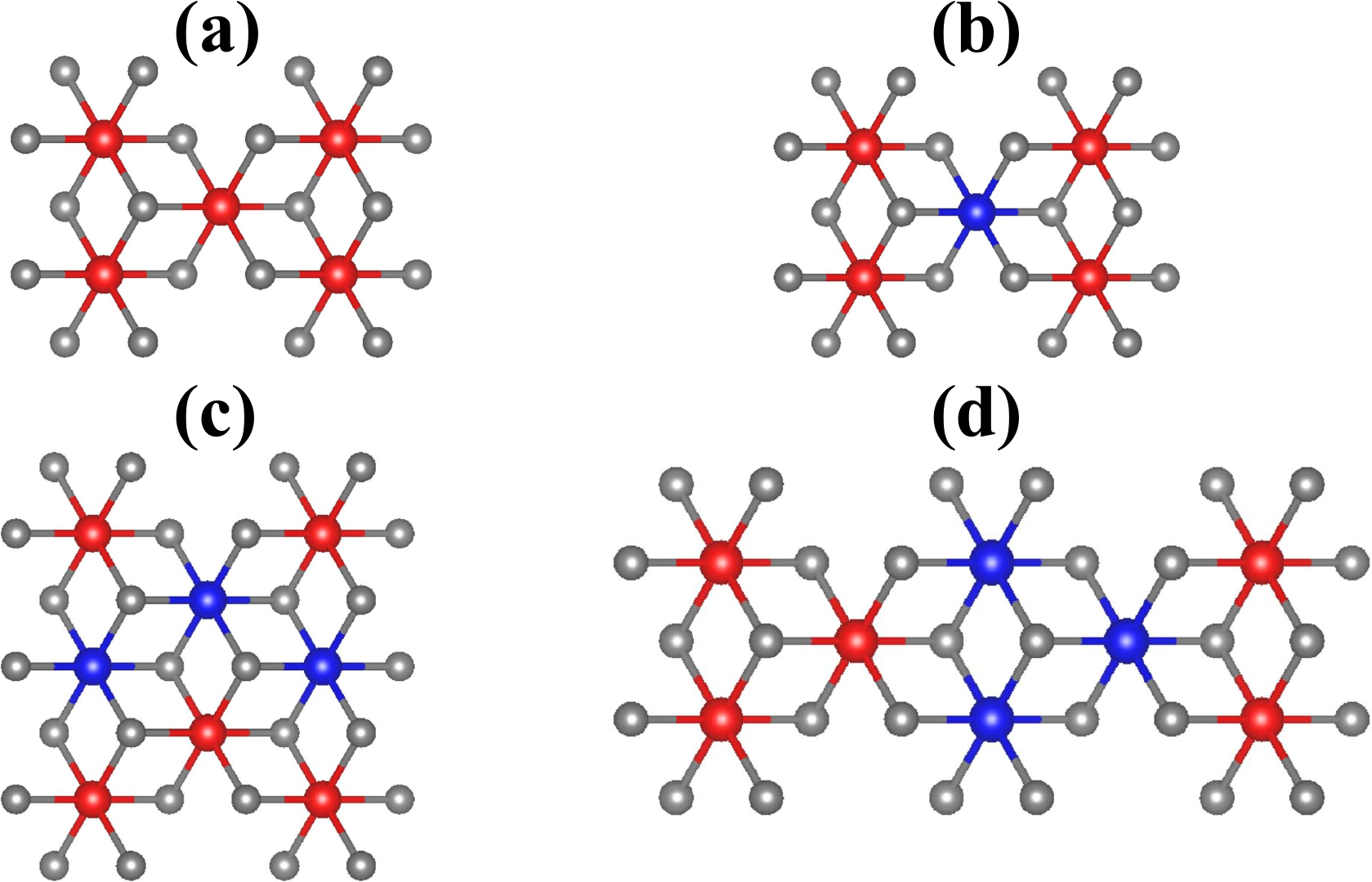}
\caption{(Color online)Four magnetic configurations, including FM (a), AFM1 (b), AFM2 (c) and AFM3 (d) orderings. The red and blue balls represent Fe atoms with spin up and spin down, respectively, while the gray balls represent Cl atoms. }\label{mc}
\end{figure}

\begin{figure*}
    \centering
    \includegraphics[width=0.90\textwidth]{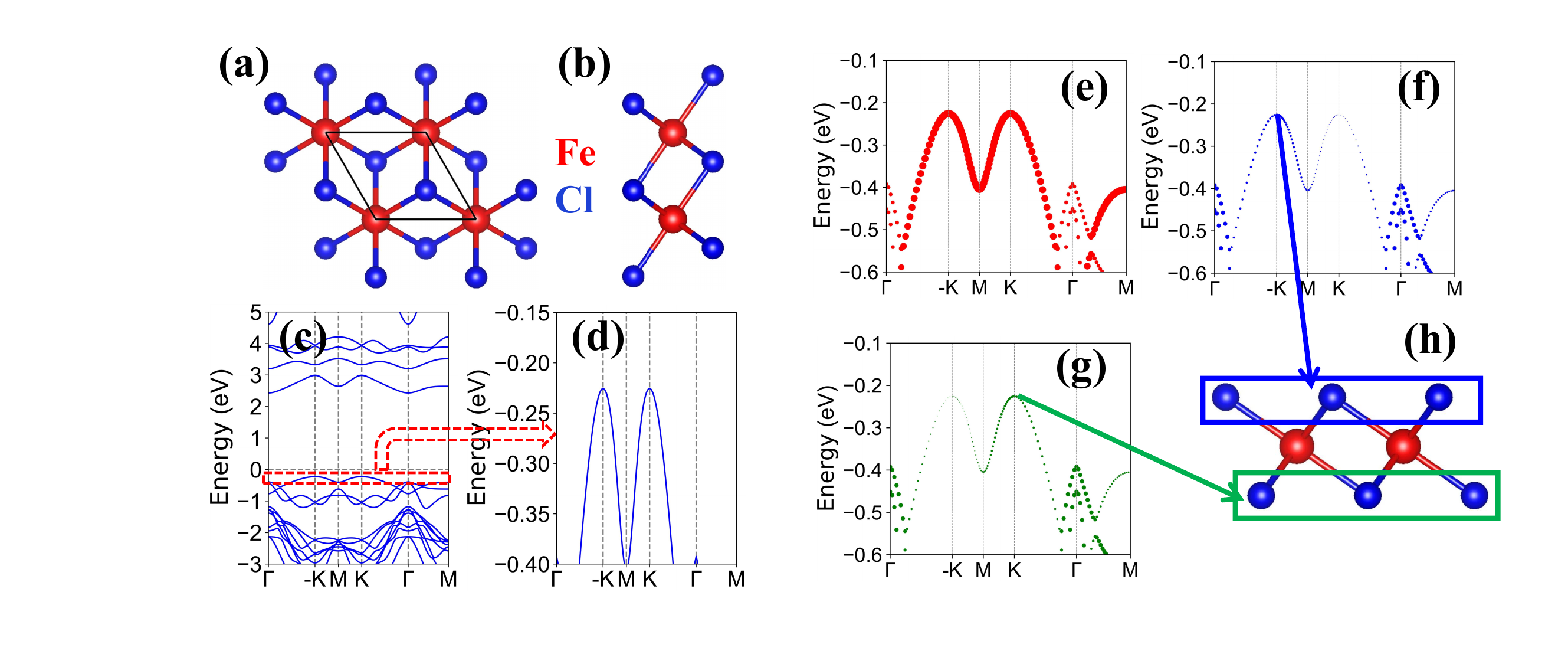}
    \caption{(Color online) For $\mathrm{FeCl_2}$, (a) and (b):top and side views of the  crystal structures; (c): the energy band structures with SOC; (d):the enlarged portion of valence bands in (c) near the Fermi energy level.  The atomic projected  energy band structure, including Fe (e), upper layer Cl (f) and lower layer Cl (g);    the schematic valley-layer coupling (f, g and h).   In (a, b, h), the red and blue balls represent the Fe and Cl atoms, respectively. In (e, f and g), the size of the radius at this point is proportional to the weight contribution.}\label{st}
\end{figure*}

\begin{figure}
    \centering
    \includegraphics[width=0.45\textwidth]{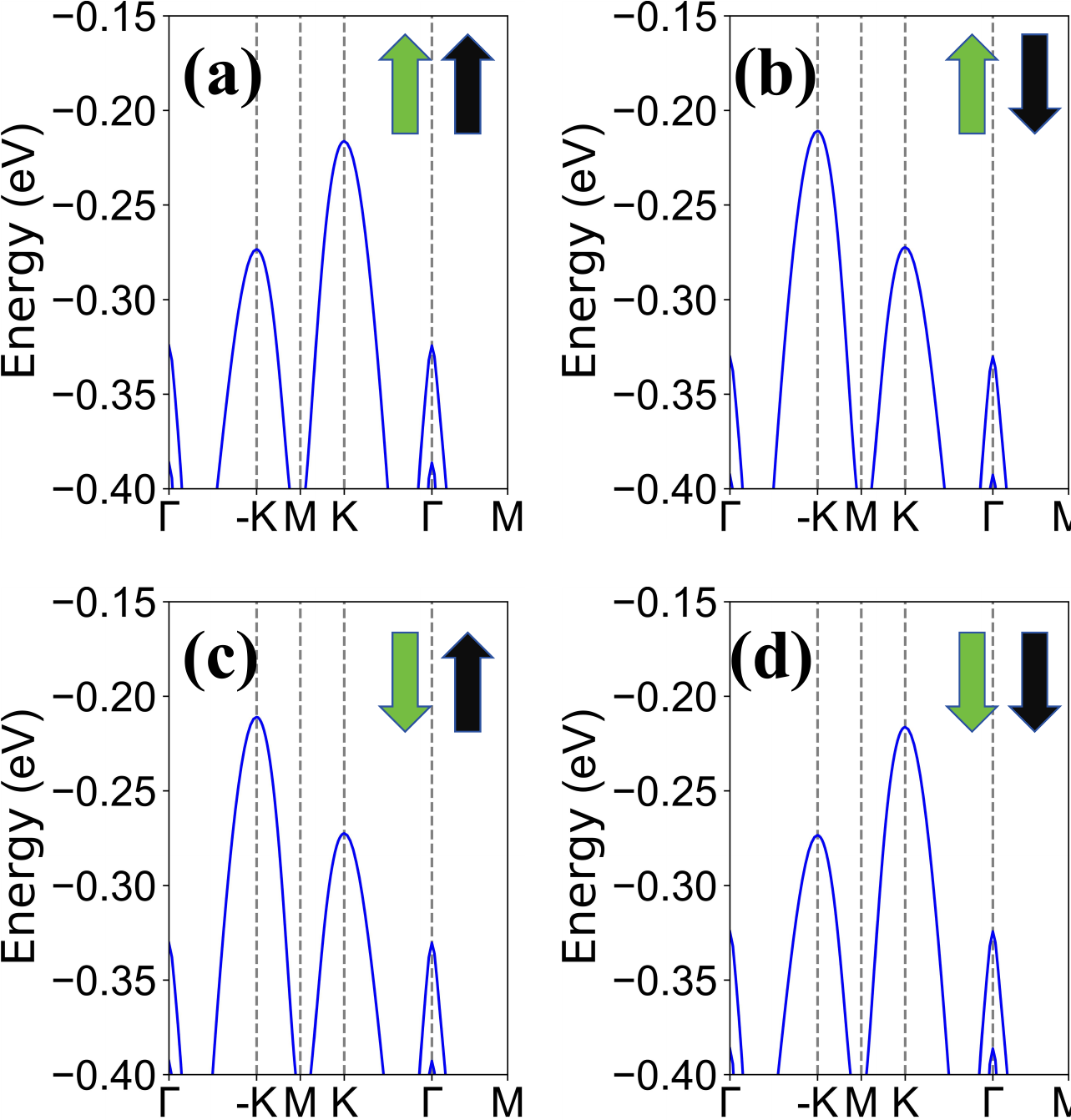}
     \caption{(Color online)For $\mathrm{FeCl_2}$, with the magnitude of the electric field strength  of 0.25 $\mathrm{V/{\AA}}$,  the enlarged portion of valence bands near the Fermi energy level with +$E$ and +$M$ (a), +$E$ and -$M$ (b), -$E$ and +$M$ (c), -$E$ and -$M$ (d).   The green and black arrows represent the direction of the electric field and the direction of magnetization, respectively.   }\label{band}
\end{figure}
\begin{figure*}
    \centering
    \includegraphics[width=0.96\textwidth]{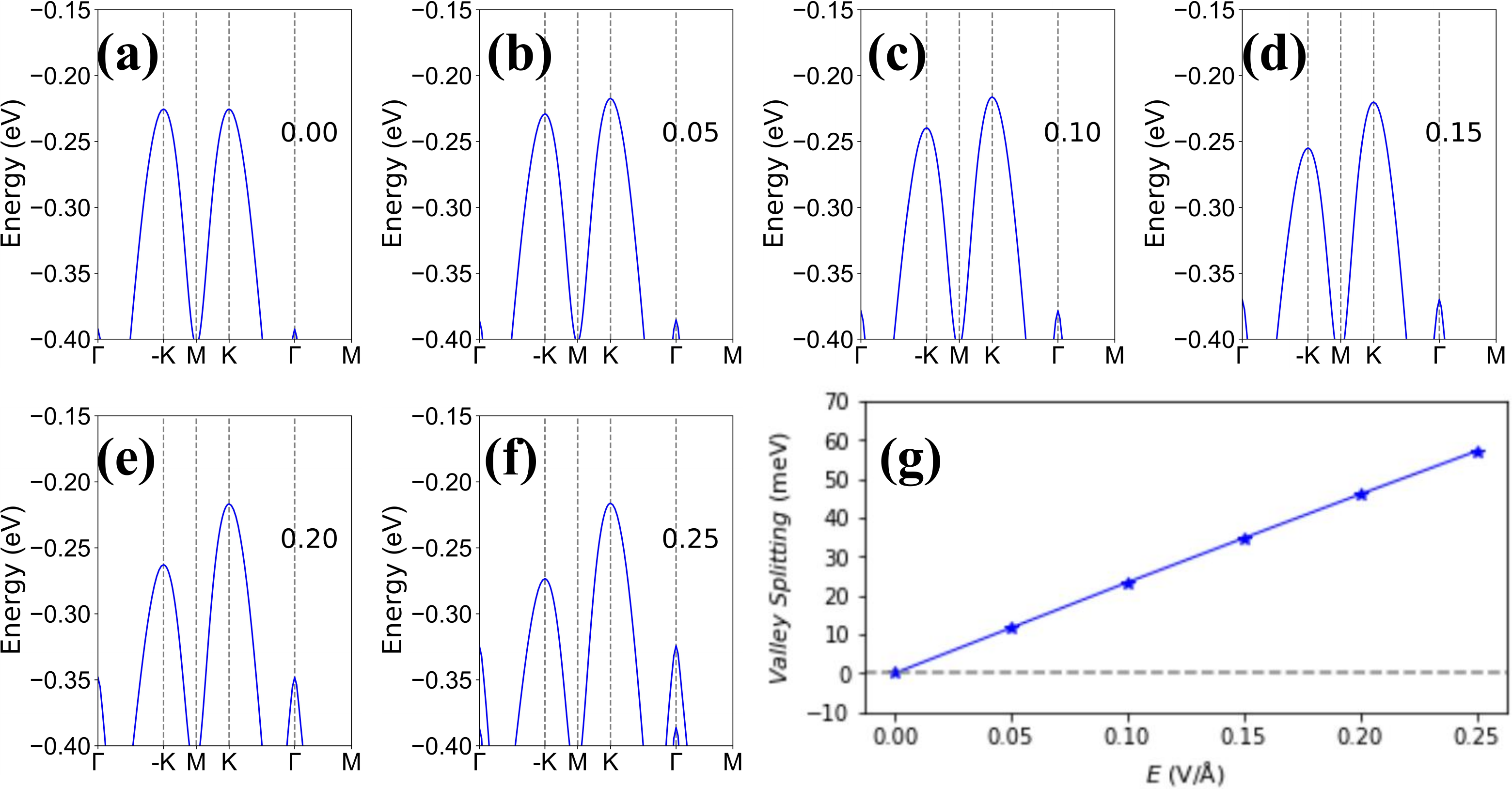}
     \caption{(Color online)For $\mathrm{FeCl_2}$, the enlarged portion of valence bands near the Fermi energy level with $E$=+0.00 (a), +0.05 (b), +0.10 (c),  +0.15 (d), +0.20 (e) and +0.25 (f)  $\mathrm{V/{\AA}}$;  the valley splitting as a function of $E$ (g).  }\label{band-1}
\end{figure*}

\textcolor[rgb]{0.00,0.00,1.00}{\textbf{Crystal structures and valley-layer coupling.---}}
Bulk $\mathrm{FeCl_2}$ possesses  a natural
layered structure\cite{v1},  and the individual layer can be
 extracted from the bulk using exfoliation techniques, which is  similar to
graphene and $\mathrm{MoS_2}$.  Monolayer $\mathrm{FeCl_2}$ has been experimentally
synthesized\cite{v2,v3}.    The top and side views of the  crystal structures of monolayer  $\mathrm{FeCl_2}$ are shown in \autoref{st} (a) and (b),  which  is composed of
three atomic planes: a Fe layer sandwiched between two Cl layers. $\mathrm{FeCl_2}$ monolayer  crystallizes in the  $P\bar{3}m1$ (No.~164),  possessing lattice $P$ symmetry.  As plotted in \autoref{mc}, four magnetic configurations, including FM, AFM1, AFM2 and AFM3 orderings,  are considered to determine magnetic  ground state.
The energies of AFM1, AFM2, and AFM3 are 118 meV, 1.266 eV, and  42 meV higher than that of FM, respectively, which implies that the FM ordering is the ground state.
The energy of AFM2 is much higher than that of FM because the magnetic order of AFM2 converges to a non-magnetic solution.
The optimized  equilibrium lattice constants are $a$=$b$=3.53 $\mathrm{{\AA}}$ with FM ordering by using GGA, which agree well with previous calculated results\cite{fecl}.  The direction of magnetization significantly influences valley-layer coupling, and  $\mathrm{FeCl_2}$ with only out-of-plane magnetization can exhibit valley-layer coupling (see below).
The magnetic orientation can be determined by MAE, and the calculated value is 95 $\mathrm{\mu eV}$/Fe,  which  indicates  the out-of-plane easy magnetization axis of $\mathrm{FeCl_2}$.

By GGA+U+SOC,  the energy band structures of  $\mathrm{FeCl_2}$ with out-of-plane  magnetization are shown in \autoref{st} (c) and (d),  and  the valence band maximum (VBM) is located at the -K/K valley.  It is clearly seen that there is no valley polarization between the -K and K valleys due to $P$ symmetry.
Nevertheless, $\mathrm{FeCl_2}$ can exhibit weak valley-layer coupling, which provides the basis for electric field-induced valley polarization.
Based on the band structure with atomic projections in \autoref{st} (e, f, g and h), the Fe atoms contribute equally to the -K and K valleys, while the upper/lower Cl atoms mainly contribute to the -K/K valleys,  which gives rise to the so-called valley-layer coupling. Compared to the existing valley-layer coupling\cite{k11,k12,k13}, this is a weak valley-layer coupling in  $\mathrm{FeCl_2}$, because the atomic contributions to the two valleys mainly come from Fe atoms in the same layer, rather than from atoms in different layers.  Under the same interlayer distance, a larger electric field is required to achieve the same valley splitting for weak valley-layer coupling. Calculated results show that  K and -K valleys are mainly from Fe-$d_{x^2-y^2}$+$d_{xy}$ (see FIG.S1\cite{bc}), which provides the fundamental conditions for electric field-induced valley polarization.
Finally, it should be pointed out that out-of-plane magnetization is required to achieve valley-layer coupling. According to FIG.S2\cite{bc}, in the case of in-plane magnetization, both the Fe  atoms and  the upper Cl and lower Cl atoms contribute equally to the -K and K valleys, and there is no valley-layer coupling.
In other words, valley polarization can only be induced in the case of out-of-plane magnetization when an additional out-of-plane electric field is applied.

\textcolor[rgb]{0.00,0.00,1.00}{\textbf{Electric field-induced valley polarization.---}}
The electric field  can generate a layer-dependent electrostatic potential, which can induce valley polarization in 2D valley-layer coupling materials\cite{k11,k12,k13}.
For $\mathrm{FeCl_2}$, with the magnitude of the electric field strength  of 0.25 $\mathrm{V/{\AA}}$,  the enlarged portion of valence bands near the Fermi energy level with +$E$ and +$M$, +$E$ and -$M$, -$E$ and +$M$, -$E$ and -$M$ are plotted in \autoref{band}, where  the+$E$/-$E$ represents the electric field direction along the positive/negative $z$-axis, and the +$M$/-$M$ indicates the magnetization direction along the positive/negative $z$-axis.
It is clearly seen that all four cases produce the same valley splitting of 57.2 meV ($\Delta E=E_{K}-E_{-K}$), and the valley polarization can be reversed by inverting the direction of magnetization or the electric field. Reversing the direction of either the magnetization or the electric field alone can invert the valley polarization (\autoref{band}(a)$\rightarrow$\autoref{band}(b) or \autoref{band}(a)$\rightarrow$\autoref{band}(c)), while reversing both directions simultaneously leaves the valley polarization unchanged (\autoref{band}(a)$\rightarrow$\autoref{band}(d)).
If we use binary values of +1 and -1 to represent the positive and negative directions of $E$  and $M$, then the K-valley polarization occurs when the product of $E$  and $M$ is +1, and the -K-valley polarization occurs when the product is -1.  For 2H-$\mathrm{FeCl_2}$, only the reversal of the magnetization direction can be used to flip the valley polarization\cite{v4}.  For our investigated  1T-$\mathrm{FeCl_2}$, the coupling of the electric field and magnetization provides more flexibility for valley control.

Next, we consider the effect of the electric field amplitude on the magnitude of valley splitting.
 The magnetic ground state of $\mathrm{FeCl_2}$ under the applied electric field is determined by comparing the energies of different magnetic configurations.
As plotted in FIG.S3\cite{bc},  within considering $E$ range, $\mathrm{FeCl_2}$  monolayer always has FM ordering.
An important factor for the generating valley-layer coupling is out-of-plane magnetization. According to FIG.S4\cite{bc}, the positive MAE ensures that $\mathrm{FeCl_2}$ remains out-of-plane magnetization within the considered range of electric field.
For $\mathrm{FeCl_2}$, the enlarged portion of valence bands near the Fermi energy level at the representative electric field $E$ and  the valley splitting as a function of $E$ are shown in \autoref{band-1}.  The calculated results show that the valley splitting increases linearly with the increase of $E$.
For strong valley-layer coupling, the valley splitting can be approximately calculated using   $eEd$, where $e$  denotes the electron charge,  and the $d$  is the  interlayer distance.  For weak valley-layer coupling, a similar expression can also be used for the valley splitting calculation, but it needs to be multiplied by a reduction factor $\alpha$.  For  $\mathrm{FeCl_2}$,  $\alpha eEd$ can be used to estimate the valley splitting, where  $\alpha$=0.081 and $d$=2.82 $\mathrm{{\AA}}$.
For in-plane magnetization, valley polarization will not exist even with the application of an electric field (see FIG.S5\cite{bc}), due to the lack of valley-layer coupling.

If the two Cl layers are not equivalent caused by electric field, the $P$ and  horizontal mirror symmetries of $\mathrm{FeCl_2}$ are broken,   which will crystallize in the  $P3m1$ (No.~156).  When the group symmetry of the wave vector at K/-K valley becomes $C_{3h}$,
  for $d_{x^2-y^2}$/$d_{xy}$-dominated -K/K valley, the valley splitting can be written as : $\Delta E_V=4\alpha cos\theta$\cite{v5} ($\alpha$ is SOC-related parameter, and the $\theta$=0/90$^{\circ}$ means out-of-plane/in-plane direction.). According to FIG.S1\cite{bc}, it is found that  the  -K and K valleys of valence bands in $\mathrm{FeCl_2}$ are mainly from $d_{x^2-y^2}$/$d_{xy}$ orbitals. When the magnetization direction is out-of-plane/in-plane,  the valley splitting of $\mathrm{FeCl_2}$ will be 4$\alpha$/0.

\begin{figure*}[t]
    \centering
    \includegraphics[width=0.85\textwidth]{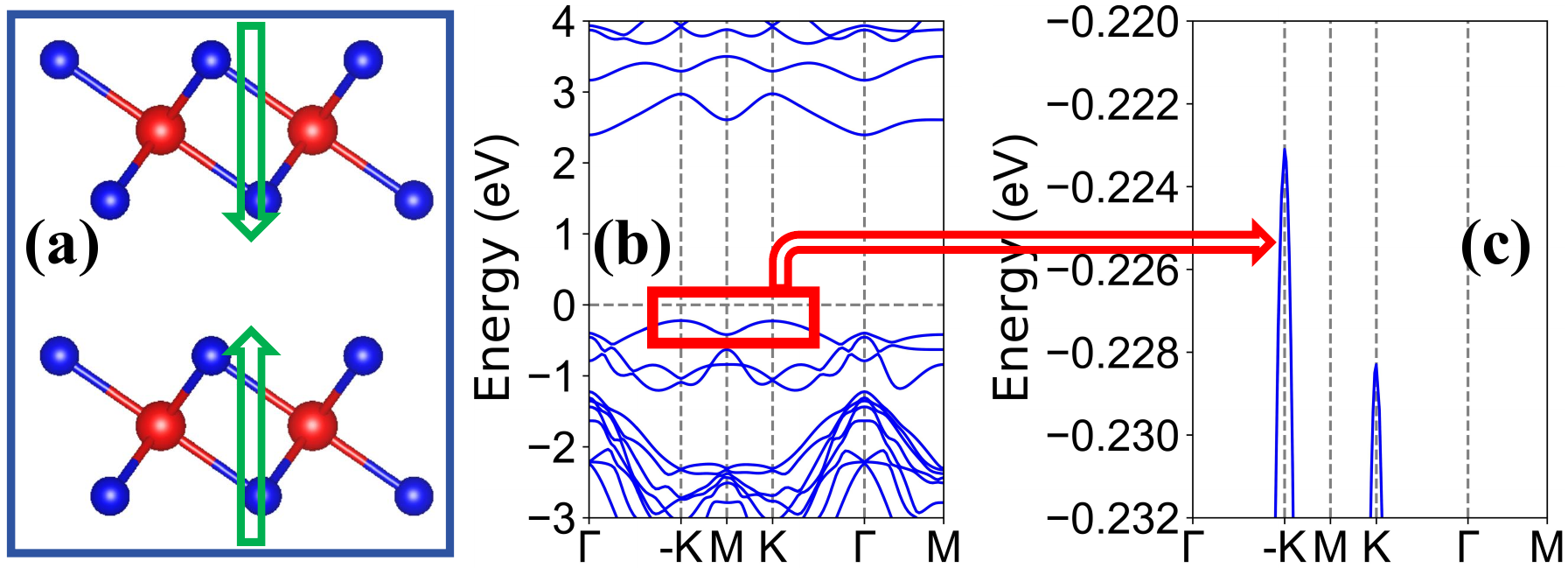}
     \caption{(Color online)For bilayer  $\mathrm{FeCl_2}$ with AA stacking, (a): side view of the  crystal structures; (b): the energy band structures with SOC; (c):the enlarged portion of valence bands in (b) near the Fermi energy level.  In (a), the red and blue balls represent the Fe and Cl atoms, and the green arrows represent the built-in electric field.  }\label{bi}
\end{figure*}

\textcolor[rgb]{0.00,0.00,1.00}{\textbf{Bilayer stacking-induced valley polarization.---}}
When two $\mathrm{FeCl_2}$ are stacked together, one can be considered as the substrate, and the other will experience an electric field, as shown in \autoref{bi} (a).  Thus, each $\mathrm{FeCl_2}$ will generate valley polarization. If the two $\mathrm{FeCl_2}$ are stacked and magnetically coupled appropriately, the overall bilayer will possess valley polarization.

Two types of stacking, namely AA (\autoref{bi}) and AB (see FIG.S6\cite{bc}), are considered,  both of which possess $P\bar{3}m1$ (No.~164)  space group with lattice $P$ symmetry.
To determine the magnetic ground state of these bilayers, the intralayer FM and interlayer AFM  (FM-AFM), and intralayer FM and interlayer FM  (FM-FM) configurations are considered. For both AA and AB stacking, the FM-AFM  magnetic configurations are the ground state, and their energies are 3.3 and 2.7 meV lower than those of the FM-FM cases.  After comparing the energies of the FM-AFM magnetic configurations for AA and AB stacking, it is found that the energy of AA stacking is 2.8 meV lower than that of AB stacking. This confirms that AA stacking is the ground state, which is consistent with existing experimental and computational results\cite{v3}.

For AA stacking, the energy band structures are plotted in \autoref{bi} (b) and (c). Due to $PT$ symmetry,   bilayer  $\mathrm{FeCl_2}$ is a $PT$-antiferromagnet, possessing the characteristic of spin degeneracy.  It is clearly observed that there is a valley polarization between the -K and K valleys, with a valley splitting of about 6 meV. This finding challenges our original proposal that, to  achieve valley polarization in a $PT$-antiferromagnetic bilayer, the fundamental building block should inherently possesses valley polarization\cite{v6}. For the AB stacking, this valley polarization also exists with very small valley splitting of 0.5 meV (see FIG.S6\cite{bc}).
  Compared to the upper layer $\mathrm{FeCl_2}$, the magnetization and electric field directions of the lower layer $\mathrm{FeCl_2}$ are simultaneously inverted. Therefore, based on \autoref{band}, the upper and lower layers  have the same valley polarization. The bilayer $\mathrm{FeCl_2}$ possesses horizontal mirror symmetry, so the energy bands of the upper and lower layers will coincide with each other.

\textcolor[rgb]{0.00,0.00,1.00}{\textbf{Discussion and conclusion.---}}
Experimentally, both monolayer and bilayer  $\mathrm{FeCl_2}$ have been synthesized\cite{v2,v3}.
Recently, a significant advancement has been made in the field of 2D materials, where an intense electric field exceeding 0.4 $\mathrm{V/{\AA}}$ has been achieved through dual ionic gating\cite{zg7}. This breakthrough provides a promising avenue for realizing large valley splitting in monolayer $\mathrm{FeCl_2}$.
 Very recently,  switching the spin polarization of the conduction band on and off has been experimentally realized in bilayer  $\mathrm{CrPS_4}$ by a perpendicular electric field\cite{nn}, and the underlying mechanism mirrors the process of electric field-induced valley polarization in materials with valley-layer coupling.  Essentially, both mechanisms involve out-of-plane electric-field-induced layer-dependent potential, which offsets the electronic states of the two layers.   These existing experiments lay the foundation for the control of valley polarization in  $\mathrm{FeCl_2}$ using electric field.

In summary, first-principles calculations reveal weak valley-layer coupling in centrosymmetric $\mathrm{FeCl_2}$ monolayer. Unlike strongly coupled monolayers, $\mathrm{FeCl_2}$ requires an extremely strong electric field to achieve comparable valley splitting. Valley-layer coupling exists only when the magnetization is out-of-plane, and it is under this condition that the electric field can induce valley polarization. Valley polarization switching can be controlled by manipulating the directions of magnetization and electric field: reversing one direction switches the polarization, while reversing both simultaneously leaves it unchanged. Additionally, bilayer $\mathrm{FeCl_2}$, as a $PT$-antiferromagnet, can spontaneously achieve valley polarization without an external electric field.
Our work offers an intuitive understanding of electric-field-induced valley polarization in  $\mathrm{FeCl_2}$, and  we propose a strategy for inducing valley polarization through bilayer stacking.

\begin{acknowledgments}
This work is supported by Natural Science Basis Research Plan in Shaanxi Province of China  (2025JC-YBMS-008). We are grateful to Shanxi Supercomputing Center of China, and the calculations were performed on TianHe-2.
\end{acknowledgments}

\end{document}